\documentclass[twocolumn,showpacs,preprintnumbers,amsmath,amssymb]{revtex4}

\usepackage{graphicx}% Include figure files
\usepackage{dcolumn}% Align table columns on decimal point
\usepackage{bm}% bold math
\usepackage{braket}
\usepackage{psfrag}
\usepackage{epsfig}
\usepackage{color}
\begin{document}

\title{Alpha Decay Hindrance Factors: A Probe of Mean Field Wave Functions}

\author{D. Karlgren} \altaffiliation[Also at]{Instituut voor Kern- en
 Stralingsfysica, University of Leuven, Celestijnlaan 200 D, B-3001
 Leuven, Belgium}
 \author{R .J. Liotta} 
\author{R. Wyss}
 \affiliation{Department of Physics, Royal Institute of Technology,
 104 05 Stockholm, Sweden} 
\author{M. Huyse} 
\author{K. Van de Vel}
 \author{P. Van Duppen}
\affiliation{Instituut voor Kern- en
Stralingsfysica, University of Leuven, Celestijnlaan 200 D, B-3001
Leuven, Belgium}

\date{\today} \begin{abstract}
A simple model to calculate $\alpha$-decay
Hindrance Factors is presented. Using deformation values obtained from
PES calculations as the only input, Hindrance Factors for the
$\alpha$-decay of Rn- and Po-isotopes are calculated. It is found that
the intrinsic structure around the Fermi surface determined by the
deformed mean field plays an important role in determining the
hindrance of $\alpha$-decay. The fair agreement between experimental
and theoretical Hindrance Factors suggest that the wave function
obtained from the energy minima of the PES calculations contains an
important part of the correlations that play a role for the
$\alpha$-decay. The calculated HF that emerges from these calculations render
a different interpretation than the commonly assumed 
n-particle n-hole picture.
\end{abstract}

 \maketitle

\section{\label{sec:level1}INTRODUCTION:\protect}

The recent experimental exploration of the neutron deficient Pb-region
has given new insight into the low lying excitations of atomic
nuclei\cite{wau941}. In particular, the discovery of the triplet spin $0^+$
states as the lowest excitations in $^{186}$Pb reveals the coupling of
single particle motion to the collective shape degree of freedom in a
striking manner\cite{and00}.  The quantitative description of these
excitations as well as their isotopic and isobaric dependence is a
real challenge to nuclear structure models.  The balance of prolate
and oblate shapes e.g. depends crucially on the iso-spin dependence
of the spin orbit interaction as well as the particular shape of the
nuclear potential\cite{taj01}.  

This paper will focus on the $\alpha$ decay as a tool to understand
the microscopic structure related to the different shapes in the
Pb-region within the frame work of the mean field model. To calculate
the absolute values of the $\alpha$ decay rate is a problem which is
still not fully solved. Instead, we will investigate the ratio of the
Hindrance Factors of the $\alpha$ decay to different states. In this
manner, only parts of the absolute decay width needs to be taken into
account making the problem more tractable. We limit the discussion to
$\alpha$ decay of the $0^+$ ground state of even-even nuclei to the
$0^+$ states in the daughter nuclei; $\Delta L = 0$ transitions.

The Hindrance Factor is the ratio between the reduced $\alpha$ decay
widths, labeled $\delta^2$, of the ground state-to-ground state decay
and the reduced $ \alpha$ decay width of the decay to an excited state in
the daughter nucleus \cite{ras59}. The experimental reduced
$\alpha$ decay widths are calculated using a spherical potential
barrier. In principle, using a deformed potential barrier could
influence the calculated $\delta^2$-value, but the effect is expected to be
very small for the zero angular momentum decays treated in this
work. The Hindrance Factor is independent of the $\alpha$-decay energy
and can therefore be used to compare the intrinsic mechanisms that
induce the $\alpha$-decay process. This is an important property in
the study of shape coexisting nuclei, since the decay between states
with similar single particle structure is thought to take place more
easily than that between states with different single particle
structure. In the present work the single particle structure is
determined by the deformation parameters of the universal parameter
Woods-Saxon potential, see ref. \cite{naz85}. A study of the Hindrance
Factors can therefore help in establishing the deformation of the
ground state and the excited states of the daughter nucleus if the
deformation of the decaying nucleus is known or vice versa.

Earlier work on Hindrance Factors have used a two-level mixing model
based on the spherical shell model \cite{kh87}. In this model the
intrinsic decay rates have been related to each other through
experimentally measured Hindrance Factors \cite{dup90,dup00}. A
microscopic model based on RPA calculations has been presented in ref.
\cite{del96}. In the present work we start from the mean field wave
function as obtained from Potential Energy Surface (PES)
calculations. In this way, the deformation of each state is derived
microscopically and is not a free parameter of the calculations.

Section 1 of this paper gives an introduction to the subject of
Hindrance Factors. Section 2 describes the model used in our
calculations. In section 3 we show the dependency of Hindrance Factors
on properties such as deformation, pairing correlations and particle
number. Section 4 is a comparison between the 
$nparticle-nhole~~(np-nh)$ spherical shell
model picture and the mean-field approach applied in our model. In
section 5 we present and discuss results from calculations using the
model described in section 2.
\section{MODEL:\protect}
The $\alpha$-particle decay width can be factorized into two parts
\cite{lov98},
\begin{equation}
\Gamma(\bar{R})=P(\bar{R})\frac{\hbar^2\bar{R}}{2M}|\mathcal{F}(\bar{R})|^2,
\end{equation}
where $P(\bar{R})$ is the Coulomb penetrability at the distance $R$
between the center of mass and the $\alpha$-particle, and
$\frac{\hbar^2\bar{R}}{2M}|\mathcal{F}(\bar{R})|^2$ is the reduced
decay width \cite{ras59}. Here $\mathcal{F}(\bar{R})$ is the $\alpha$
cluster formation amplitude.

The Hindrance Factor is given by the ratio of the absolute value
squared of the $\alpha$-particle formation amplitudes for the decays
from the ground state of the mother nucleus to the ground state (gs)
and the excited state (es) of the daughter nucleus. This can be
written as
\begin{equation}
\label{equation:1}
HF=\sum_l\frac{|\mathcal{F}_l(\bar{R};gs\rightarrow gs)|^2}
{|\mathcal{F}^{'}_l(\bar{R};gs\rightarrow es)|^2},
\end{equation}
where the ground state to ground state decay is taken to be non
hindered, i. e. the Hindrance Factor is equal to one.  Further, the
formation amplitude for a decay from state B in the mother nucleus to
the state A in the daughter nucleus can be written as \cite{lov98}
\begin{equation}
\label{equation:2}
\mathcal{F}_l(\bar{R})=\int[\Phi_{\alpha}(\xi_{\alpha})\Psi_A(\xi_A)
Y_l(\bar{R})]^*\Psi_B(\xi_B)d\xi_{\alpha}d\xi_A,
\end{equation}
If we rewrite the equation \ref{equation:2} in Fock space we obtain
\begin{equation}
\label{equation:3}
\mathcal{F}_l(\bar{R})=\braket{\Phi_{\alpha};\Psi_{A}|Y^*_l(\bar{R})|\Psi_{B}}.
\end{equation}
By using Thouless theorem, see \cite{rs80}, it is possible to connect
two HFB vacuum states with different deformations. Therefore the
mother nucleus can be expanded in terms of two pair excitations of the
daughter nucleus, although the deformation of the two states may
differ. For a more thorough investigation of the problem, see
ref. \cite{del98}. An expression for the mother nucleus as a four
particle excitation of the daughter nucleus is obtained,
\begin{equation}
\label{equation:4}
\ket{\Psi_B}=\sum_{kl}B(k)B(l)c^+_lc^+_{\bar{l}}c^+_kc^+_{\bar{k}}
\ket{\Psi_A},
\end{equation}
where $k,\bar{k}$ is a pair of protons and $l,\bar{l}$ is a pair of
neutrons. The factors $B(k)$ and $B(l)$ are related to the pairing
densities
\begin{equation}
\label{equation:5}
B(k)=\braket{\Psi_B|c^+_kc^+_{\bar{k}}|\Psi_A},
\end{equation}
and can readily be evaluated using the Onishi formula
\cite{rs80}, yielding the expression
\begin{equation}
\label{equation:6}
\braket{\Psi_B|c^+_kc^+_{\bar{k}}|\Psi_A}=\braket{\Psi_B|\Psi_A}
(-U^*_AU^{T^{-1}}V^T_B)_{k\bar{k}},
\end{equation}
where 
\begin{equation}
\label{equation:7}  
U=(U_A^+U_B+V_A^+V_B),
\end{equation}
and the $U_i$ and $V_i$ are the matrices of the Bogoliubov transform
for the daughter and mother nuclei. This gives a total expression for
the formation amplitudes as
\begin{eqnarray}
\label{equation:8}
\mathcal{F}_l(\bar{R})=\braket{\Phi_{\alpha};\Psi_{A}|Y_l(\bar{R})|
\sum_{kl}B(k)B(l)c^+_lc^+_{\bar{l}}c^+_kc^+_{\bar{k}};\Psi_A}
\nonumber \\
=\braket{\Psi_{A}|\Psi_A}\sum_{kl}B(k)B(l)\braket{\Phi_{\alpha}
|Y_l(\bar{R})|c^+_lc^+_{\bar{l}}
c^+_kc^+_{\bar{k}}}.
\end{eqnarray}
The integral can be solved analytically by expressing the wave function
of the $\alpha$-particle in terms of Harmonic Oscillator wave functions
and using the Moshinsky brackets \cite{mos59} to transform to a center
of mass system. However it turns out \cite{pog69} that for the special
case of an $\alpha$-particle with angular momentum 0 all terms
$\Lambda^{\Omega{\pi}-\Omega{\pi}\Omega{\nu}-
\Omega{\nu}}_{L_{\alpha}M_{\alpha}}$ of the integral
\begin{equation}
\label{equation:9}
\Lambda^{l\bar{l}k\bar{k}}_{00}=\braket{\Phi_{\alpha}
|Y_0(\bar{R})|c^+_lc^+_{\bar{l}}
c^+_kc^+_{\bar{k}}},
\end{equation}
are positive and can as a first approximation be replaced by an average
value $\Lambda^{avr}_{00}$. If we now evaluate the expression
\ref{equation:2} for the decays between two $0^+$ states we get
\begin{eqnarray}
\label{equation:10}
HF=\frac{\mathcal{F}(\bar{R};0_1^+\rightarrow 0_1^+)}
{\mathcal{F}(\bar{R};0_1^+\rightarrow 0_2^+)}=
\frac{\Lambda^{avr}_{00}(\bar{R})\sum_{kl}B(k)B(l)}
{\Lambda^{avr}_{00}(\bar{R})\sum_{k'l'}B(k')B(l')}\nonumber \\
=\frac{\sum_{kl}B(k)B(l)}
{\sum_{k'l'}B(k')B(l')},
\end{eqnarray}
where the $0^+_1$ is the ground state and $0^+_2$ is the excited $0^+$
state. The physical meaning of equation \ref{equation:10} is that the
pair transfer amplitude for protons times that for neutrons is the
essential quantity that determines the ease with which the mother
nucleus decays into a corresponding state of the daughter nucleus.

\section{DEPENDENCY OF THE HINDRANCE-FACTOR ON PAIRING AND DEFORMATION:\protect}

The deformation affects the Hindrance Factor in two ways. First, the
degeneracy of levels at spherical shape is lifted. This means that two
nuclei with different shapes have the levels rearranged and different
levels become occupied at the Fermi surface. Without pairing
interaction this would mean that the overlap between two wave
functions describing a nucleus at different shapes becomes zero as
soon as we encounter a level crossing.

\begin{figure}[h]
\begin{center}
\resizebox{0.48\textwidth}{!}{\input{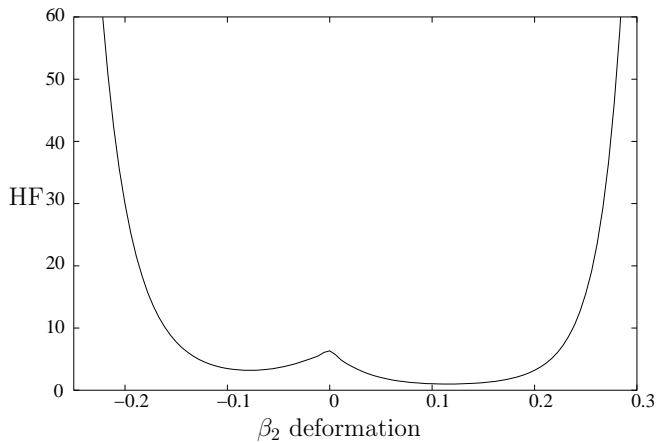}}
\caption{\label{fig:betaplot} Hindrance Factor as a function of
$\beta_2$ deformation of ${^{190}Pb}$ for the decay ${^{194}Po}
\rightarrow {^{190}Pb}$. The $\beta_2$ deformation of ${^{194}Pb}$ is
set to 0.1 and both nuclei are assumed to have the triaxiality
parameter $\gamma=0$. The result is normalized so that the HF of the
most favorable decay is set to one. Note that the minimum occurs at
$\beta_2=0.12$ in the daughter nucleus for this specific decay. The
increase of hindrance for the spherical shape is a consequence of the
decrease in pairing correlations due to the 82 shell gap.}

\end{center}
\end{figure}

\begin{figure}[h]
\resizebox{0.48\textwidth}{!}{\input{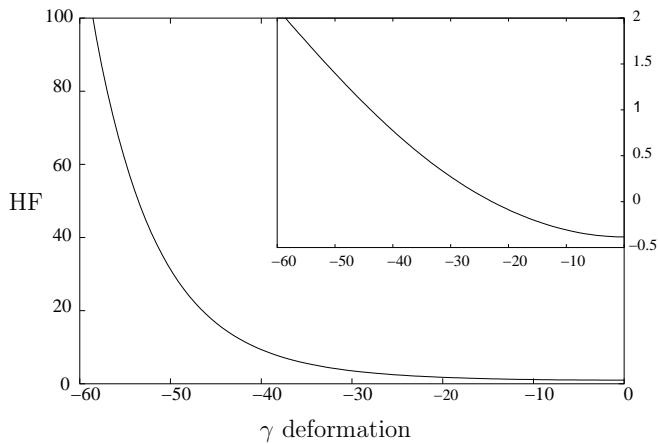}}
\caption{Hindrance Factor as a function of $\gamma$ deformation for
the decay ${^{194}Po} \rightarrow {^{190}Pb}$. The $\gamma$
deformation of ${^{194}Po}$ is set to 0, both nuclei are set to have
$\beta_2=0.2$, The result is normalized so that the HF of the most
favorable decay is set to one. The inset shows the same plot in
logarithmic scale.}
\label{fig:gammaplot}
\end{figure}
Secondly, the spherical components building up each single particle
level change smoothly with deformation. Therefore the hindrance
factors are not constant even in absence of level crossings.
 
The pairing interaction results in a smoothing of the differences in
level structure for nuclei with different shapes. This effect only
occurs in the vicinity of the Fermi surface, in an energy window that
is determined by the pairing strength. As a large part of this work
focuses on the Pb-region we have the further complication of the large
proton energy gap between proton numbers 82 and 84. The energy gap
does not allow for a non trivial solution to the BCS-equation. Since
the vanishing of the BCS solution is non physical and in order to
mimic the correlations that are indeed present in the ground state
wave function we have increased the pairing strength by 15\% which
guarantees a non trivial solution. The odd even mass difference we
obtain are well within the experimental ones, see also the discussion
in ref \cite{xu99}. This procedure is necessary in order to compare
decays to different states in a meaningful manner.

\begin{figure}[h]
\resizebox{0.4\textwidth}{!}{\includegraphics{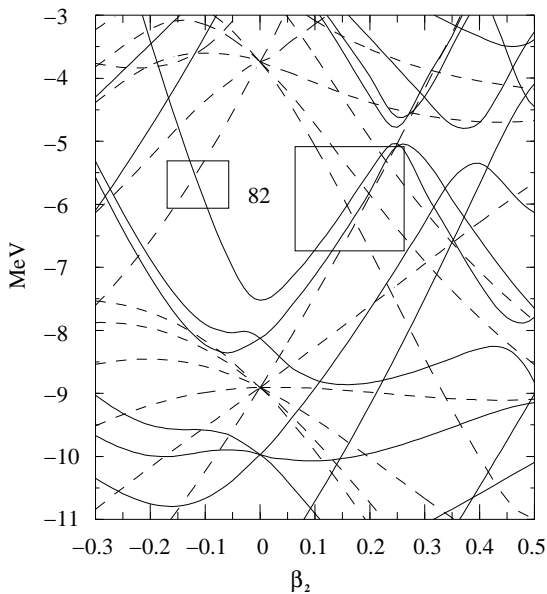}}
\caption{ Nilsson diagram for protons in the Z=82 region. 
Areas of interest from a level-crossing perspective are indicated
 both for an oblate and for a prolate deformation.}
\label{fig:pbnils}
\end{figure}

Figure \ref{fig:betaplot} shows that the HF is
only weakly dependent on deformation in an interval from
$\beta_2\approx -0.1$ to $\beta_2\approx 0.2$. By comparing to
Fig. \ref{fig:pbnils}, one notices that this is the range in
deformation, where there are no level crossings for proton number Z=82.  
However,
beyond these values in $\beta_2$ deformation, we encounter one level
crossing at oblate shape and several at prolate shape.  This is
reflected in Fig. \ref{fig:betaplot}, where the increase in HF is
considerably stronger at the prolate side compared to the oblate side.
Note the increase around spherical shape due to reduced pairing
correlations as a consequence of the Z=82 shell gap.

In fig. \ref{fig:gammaplot} the dependence on the $\gamma$ deformation
parameter is shown. Similarly to fig. \ref{fig:betaplot}, in a rather
broad range of deformation, from $\gamma=0$ to $\gamma=30$ the
increase in the HF is rather modest. However, for larger difference in $\gamma$
between mother and daughter nucleus, the HF increases steeply. In
fact, the increase in HF is exponential, as shown in the inset. In the
calculations of Fig. \ref{fig:gammaplot} we choose the mother nucleus
to be prolate. When starting from an oblate deformation for the mother
nucleus, we obtain a similar curve, but reflected at $\gamma=-30$. Our
calculations imply, that the decay from oblate to prolate is
essentially forbidden. However, we also investigated the case when the
mother has maximum triaxial deformation, i.e. a $\gamma$ value of
-30. Then the hindrance increases only with a factor of $\approx 4$
when the daughter is prolate or oblate, respectively.  Since true 
wave function has a certain spread in $\gamma$ and
$\beta_2$, the decay from prolate to oblate will proceed via the tail
of the wave function. This is of course beyond the mean field
prescription.  A soft potential in $\gamma$, resulting in a large
spread of the wave function will result in a pronounced decrease of
the HF.

\section{COMPARISON BETWEEN THE PARTICLE-HOLE 
PICTURE AND THE MEAN-FIELD PICTURE}

Shape coexistence has been described in the shell model in terms of
specific particle-hole excitations with respect to the spherical core,
see \cite{woo92}. In the Pb-region, the 2p-2h, 4p-4h or 6p-6h proton
excitation across the magical shell gap induce deformation that can
result in oblate or prolate shapes.  The spherical ground states in
the Polonium isotopes are characterized by the corresponding 2p-0h
states and the 0p-2h states correspond to the spherical Hg-isotopes.
The 2p-2h excitations are very costly in energy. On the other hand,
the energy gain due to pairing interaction and iso-scalar
quadrupole-quadrupole can partly balance the energy loss and result in
low lying deformed states.  Hence, in a qualitative fashion, one has
been able to account for the parabolic dependence as a function of
neutron number of the deformed states in the neutron deficient Pb
region as well as other regions of the nuclear chart, see e.g. the discussion
in Ref.~\cite{kh89}.

The shell model description can be linked to the deformed Nilsson (or
Hartree-Fock) scheme, by associating the particle hole excitations to
the specific level crossings, present in the Nilsson diagram, which
pin down the underlying structure of deformed states, see
\cite{kh89}. This connection is very valuable, since it enables the
comparison of the underlying structure for the two basic nuclear
structure models.  However, two aspects need to be clarified in this
respect: i) within the shell model, there is no real distinction
between oblate and prolate shapes. In the Sn-region e.g., 2p-2h
excitations corresponds to prolate shapes in the mean field model,
whereas the same excitation in the Pb region results in oblate shapes.
ii) care has to be given to the labeling of the np-nh excitations. The
2p-2h $(h_{9/2})^2-(s_{1/2})^{-2}$ 
configuration in the Pb-region e.g.,
does not agree at all with the
microscopic structure of the corresponding oblate states in the
mean-field.  In fact, the '$s_{1/2}$' hole state at oblate deformation
is dominated by the $d_{5/2}$ according to the Woods-Saxon model, and
has only a smaller contribution from the $s_{1/2}$, see ref.
\cite{and03}. The configuration mixing that is induced by deformation
is very pronounced for this particular case.  This is of course
essential for a microscopic understanding of the deformed states,
since the quadrupole moment of the $d_{5/2}$ hole is very different
from that of the $s_{1/2}$.  Realistic shell model calculations in the
Pb region with a broken proton and neutron core are far from being
feasible, but structure information from the deformed mean field can
be used as an important input to shell model calculations.

Allowed and forbidden decays in the np-nh model can be illustrated
with figure 2. of ref. \cite{dup00}. The neutrons are considered as
spectators not influencing the hindrance of the decay. Therefore the
hindrance of the $\alpha$-decay is discussed in terms of two particle
(proton) removal. An allowed decay is characterized by the removal of
two particles from the mother nucleus forming an existing state in the
daughter nucleus. The mother nucleus is in its ground state considered
to be a 2-particle state of the daughter nucleus. There are two
allowed decays from a spherical ground state; removal of the two
valence particles gives a decay to the spherical state in the daughter
nucleus; removal of two particles from the core forms a 2p-2h state in
the daughter nucleus corresponding to an oblate shape. The decay to a
prolate 4p-4h state is forbidden since it is a two step process. First
two particles have to be removed from the core, and after that another
2p-2h excitation has to take place giving a final 4p-4h state. This
means that the only allowed decay from a prolate state in the mother
nucleus is to a prolate state in the daughter nucleus. One may also
state that the $\alpha$-decay can only remove particles, not holes,
i.e. a 4p-4h configuration will remain at least a 4h configuration in
an allowed decay. The different decays are attributed with a hindrance
taken from the assumed pure (unmixed) decay between $^{198}Po$ and
$^{194}Pb$ \cite{dup00}. Once these pure states are mixed, a formula
for the Hindrance Factors can easily be obtained. The decay properties
of these mixed states can be used to determine the mixing parameters.

%I do not think that piets comment here is correct - the spherical to oblate
%should be allowed.
%For example, the spherical (proton 2p) to oblate (proton 2p-2h) transition
%should according to the particle-hole picture of ref. [8] be strongly
%hindered compared to the spherical (proton 2p) spherical (proton 0p)
%transition. In the present work it is only hindered by a factor of 2.8 (192
%Po decay) and 3.5 (190Po decay).

In the mean-field picture the different $0^+$ states are viewed as
0-quasi particle states with different deformations, described by the
standard deformation parameters $\gamma, \beta_2$ and $\beta_4$, see
ref. \cite{rs80}. From figure \ref{fig:betaplot}, one can conclude
that the underlying  single particle structure
is changing with deformation, due to the presence of distinct
level crossings at oblate and prolate shapes.
This is reflected by the slope of the hindrance factor as a function of
$\beta_2$. However, level crossings are not restricted to
spherical shapes but a generic feature of a many-body fermionic system.
In order to obtain a deeper understanding of the HF with respect to
the shell model like description, we also investigated the neutron and proton
contributions to the HF. 

\begin{figure}[h]
\resizebox{0.48\textwidth}{!}{\input{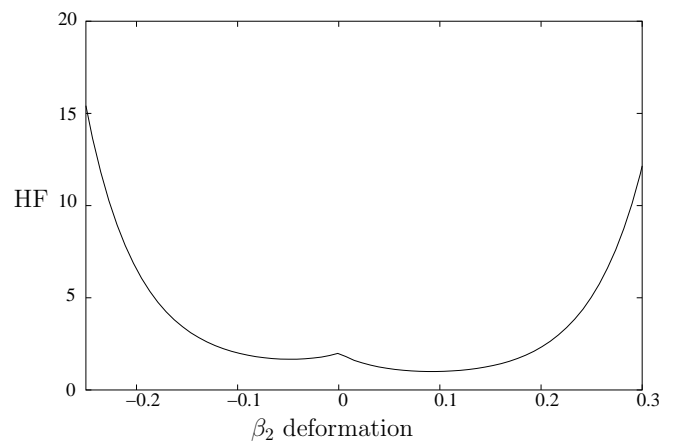}}
\caption{Contribution from neutrons to Hindrance Factor in figure
\ref{fig:betaplot}.}
\label{fig:betaplotneu}
\end{figure}
\begin{figure}[h]
\resizebox{0.48\textwidth}{!}{\input{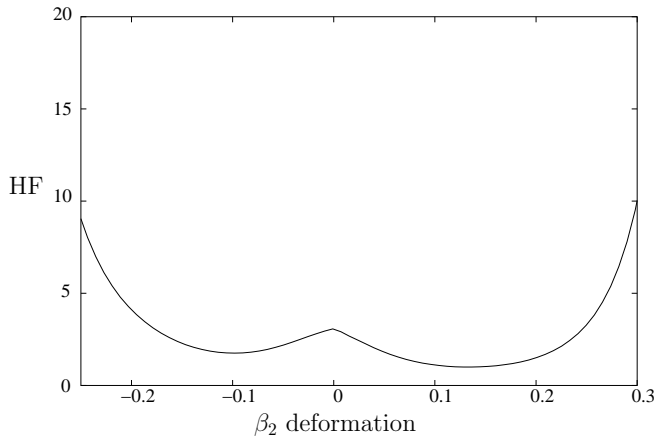}}
\caption{Contribution from protons to Hindrance Factor in figure
\ref{fig:betaplot}.}
\label{fig:betaplotprot}
\end{figure}

According to Eqs.~\ref{equation:5},\ref{equation:9}, the HF are
obtained as the product of proton and neutron contributions,
respectively.  Therefore, one can easily decompose the HF accordingly,
see Figs.~\ref{fig:betaplotneu} and \ref{fig:betaplotprot}.
Interestingly, the neutron contribution to the hindrance factor is of
similar strength as the proton part. Minor differences are seen at
spherical shape and at the largest deformations. Since the neutrons in
general are treated as spectators, see e.g. Ref. \cite{dup00} our
results suggest that this is not the case.  Therefore, in spite of the
appealing clarity of the intruder picture in terms of the np-nh
excitations with respect to the magic shell gap at Z=82, there is
little evidence that the Hindrance Factor is dominated by the change
of the proton configuration. In other words, even if the proton Fermi
level would be placed in the middle of the shell, as it is the case
for the neutrons, the calculated HF would be similar. One may
conclude, that the similarity of proton and neutron contribution to
the HF indeed indicates that this is a generic feature of the nuclear
spectrum, and not so sensitive to the details of the shell structure.

% The fact that at prolate deformation several level
%crossings appear at $\beta_2\approx 0.2$ lead to a slope that is
%clearly stronger than the one at $\beta_2\approx 0.15$ on the oblate
%side where fewer level crossings appear, see figure
%\ref{fig:pbnils}. Note also the rather flat curve for $\beta_2=-0.1$
%to $\beta_2=0.2$. It implies that in this region the $\alpha$ decay
%only modestly depends on deformation.
\begin{figure}[h]
\scalebox{0.65}{\includegraphics{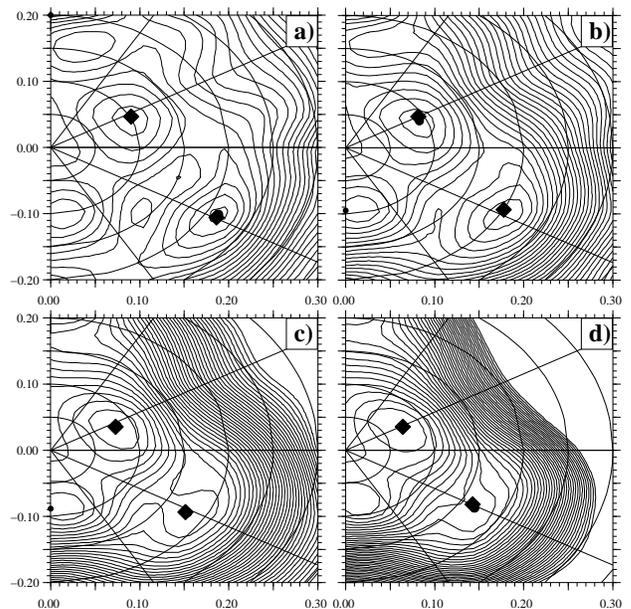}}
\caption{\label{fig:poltrs} Potential Energy Surfaces of: a)
$^{192}Po$, b)$^{194}Po$, c) $^{196}Po$ and d) $^{198}Po$, with energy
minima labeled with diamonds. The energy spacing of the contour lines
is 50 keV. The x-axis represents $\beta_2\cos(\gamma+30)$ and the
y-axis $\beta_2\sin(\gamma+30)$}
\end{figure}
\begin{figure}[h]
\scalebox{0.65}{\includegraphics{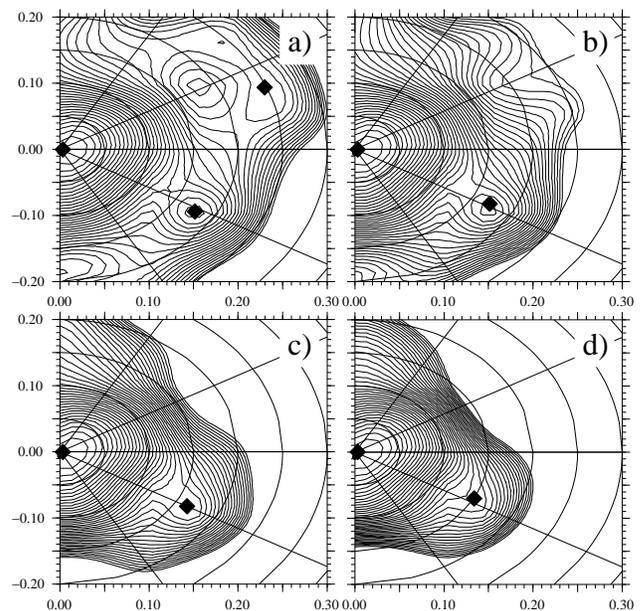}}
\caption{\label{fig:leadtrs} Potential Energy Surfaces of: a)
$^{188}Pb$, b)$^{190}Pb$, c) $^{192}Pb$ and d) $^{194}Pb$, with energy
minima marked with diamonds. The energy spacing of the contour lines
is 50 keV. The x-axis represents $\beta_2\cos(\gamma+30)$ and the
y-axis $\beta_2\sin(\gamma+30)$}
\end{figure}
\begin{figure}[h]
\scalebox{0.65}{\includegraphics{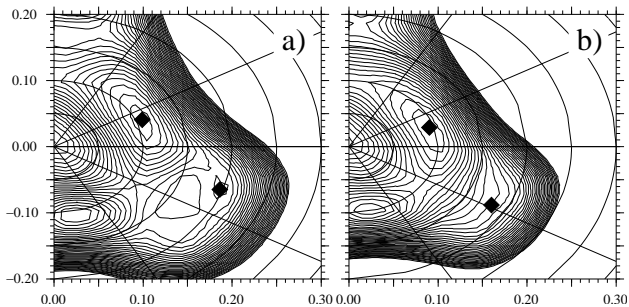}}
\caption{\label{fig:rntrs} Potential Energy Surfaces of: a)
$^{202}Rn$, b)$^{200}Rn$ with energy minima marked with diamonds. The
energy spacing of the contour lines is 50 keV. The x-axis represents
$\beta_2\cos(\gamma+30)$ and the y-axis $\beta_2\sin(\gamma+30)$}
\end{figure}

\section{RESULTS AND CONCLUSIONS}

The deformation values used for the calculations of the hindrance
factors in equation \ref{equation:10} are determined from the minima
of the PESs, see figures \ref{fig:poltrs}, \ref{fig:leadtrs},
\ref{fig:rntrs}. Extensive PES calculations of the $Po$ isotopes have
been presented earlier in ref. \cite{oro99}. We start with the decays
of selected $Po$ and $Rn$ nuclei, which are calculated to have well
separated minima. The results of the calculations are presented in
table 1. Given the simplicity of the model, the
calculations agree surprisingly well with the data. The results for
the decays of the heavier $Po$ to $Pb$ are better reproduced than that
of the $Rn$ to $Po$ decays, possibly indicating that the deformation
of the oblate states in the $Po$ isotopes is slightly underestimated
in the PES calculations. Still, the trend is nicely reproduced. For
the lightest $Po$ isotope, ${^{188}Po}$, the prolate state is calculated
to have the lowest energy, which is why the hindrance of the decay to
the excited (prolate) state in ${^{184}Pb}$ is less than one. Also for
this case, we reproduce the trend of this change, but in absolute
numbers the calculated Hindrance Factor is three orders of magnitude
off. 

\begin{table}[h]
\label{tab:results}
\caption{Calculated Hindrance Factors compared to experimental ones.}
\begin{tabular}{cccc}
  \hline Mother Nucleus & $HF_{The.}$ &$HF_{Exp.}$ \\
  \hline${^{198}Po}$ &2.82 & 3.2(5)\footnotemark[1] \\
  ${^{196}Po}$ & 2.38 & 2.6(2)\footnotemark[1] \\
  ${^{194}Po}$ & 2.41 & 1.2(2)\footnotemark[1] \\
  ${^{188}Po}$ & 1.9E-5 & 0.08(3)\footnotemark[1] \\
  ${^{202}Rn}$ & 5.46 & 19(6)\footnotemark[2] \\
  ${^{200}Rn}$ & 10.4 & 85(7)\footnotemark[3] \\
  
%  ${^{192}Po}_{sph}$ & $0_1^+$ &
\hline
\end{tabular}
\end{table}
\begin{figure}[h]
\resizebox{0.48\textwidth}{!}{\input{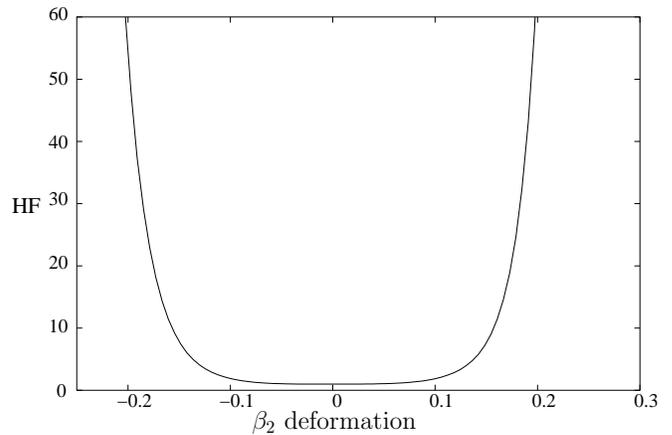}}
\caption{Hindrance Factor as a function of $\beta_2$ deformation of
${^{190}Pb}$ for the decay ${^{194}Po} \rightarrow {^{190}Pb}$. The
$\beta_2$ deformation of ${^{194}Pb}$ is set to 0.0 and both nuclei are
assumed to have the triaxiality parameter $\gamma=0$. The result is
normalized so that the HF of the most favorable decay is set to
one.}
\label{fig:betaplotspher}
\end{figure}

A remark is necessary concerning the important role played by the
small quadrupole deformation, $\beta_2\sim 0.1$, of the $Po$ isotopes,
see fig. \ref{fig:poltrs}. This slight deformation, in combination
with the reduced pairing correlations at spherical shape, is essential
to reproduce the similar strengths of the decay to the spherical and
oblate states, respectively, in the $Pb$-isotopes. If the $Po$
isotopes would be spherical in the mean field calculations the decay
from spherical to oblate would be strongly suppressed.  This is
demonstrated in Fig. \ref{fig:betaplotspher} showing that for
spherical shapes, the decay to either oblate or prolate is essentially
forbidden.  This is also at variance with the simple np-nh shell model
picture where the decay from spherical to oblate is allowed.

\footnotetext[1]{From\cite{vel03}} 
\footnotetext[2]{From\cite{wau941}}
\footnotetext[3]{From \cite{bij95}}

\begin{table}[h]
\label{tab:results2}
\caption{Calculated Hindrance Factors compared to experimental
ones. The normalization is such that the decay between the spherical
states are set to 1 as this proved to be the best way of comparing the
calculated and experimental data. The states are labeled as
1. spherical=s, 2. oblate=o, 3. prolate=p.}
\begin{tabular}{cccc}
  \hline Mother Nucleus & State in Daughter & $HF_{The.}$ 
&$HF_{Exp.}$\footnotemark[1] \\
  \hline${^{192}Po}_s$ & $0_s^+$ & 1.0 & 1.0 \\
  ${^{192}Po}_s$ & $0_o^+$ & 1.9 & 0.56(7) \\
  ${^{192}Po}_s$ & $0_p^+$ & 2.8 & $\ge$50 \\
  ${^{192}Po}_o$ & $0_s^+$ & 57 & 1.0 \\
  ${^{192}Po}_o$ & $0_o^+$ & 0.20 & 0.56(7) \\
  ${^{192}Po}_o$ & $0_p^+$ & 8.4 & $\ge$50 \\
\hline
  ${^{190}Po}_s$ & $0_s^+$ & 1.0 & 1.0 \\ 
  ${^{190}Po}_s$ & $0_o^+$ & 1.3 & 0.57(12) \\
  ${^{190}Po}_s$ & $0_p^+$ & 3.5 & 2.4(9) \\
  ${^{190}Po}_o$ & $0_s^+$ & 44 & 1.0 \\
  ${^{190}Po}_o$ & $0_o^+$ & 0.17 & 0.57(12) \\
  ${^{190}Po}_o$ & $0_p^+$ & 63 & 2.4(9) \\
  ${^{190}Po}_p$ & $0_s^+$ & 300 & 1.0 \\
  ${^{190}Po}_p$ & $0_o^+$ & 3.4 & 0.57(12) \\
  ${^{190}Po}_p$ & $0_p^+$ & 0.20 & 2.4(9) \\
\hline
\end{tabular}
\end{table}

The steep increase in hindrance for the decay of ${^{200}Rn}$ has
earlier been attributed to a phase transition, see \cite{del96}, where
the structure has changed from the decay of the heavier isotope
${^{202}Rn}$. In our calculations the increase of hindrance is an
effect of the increased deformation of the excited state in
${^{196}Po}$ compared to that of the heavier isotope
${^{198}Po}$. Since we are in a region of a level crossing, see
figures \ref{fig:betaplot} and \ref{fig:pbnils}, the calculations are
very sensitive to small deformation changes as they give a large
change in single particle structure. This is also in good agreement
with the PES calculations presented in figures \ref{fig:poltrs} and
\ref{fig:rntrs}, where only a slight difference in deformation between
the two Radon isotopes doubles the hindrance of the decay. One notices
from figures \ref{fig:gammaplot} and \ref{fig:betaplot} that only
slight changes in the equilibrium deformation are necessary in order
to reproduce experimental data.

We also calculated the Hindrance Factors for the the decays
${^{192}Po}\rightarrow{^{188}Pb}$ and
${^{190}Po}\rightarrow{^{186}Pb}$. For these decays the different
minima in the PES of both mother and daughter nuclei are less well
separated. In general, one expects the mean field to give a proper
description only when the energy minima are sufficiently deep, and the 
spreading of the wave function can be neglected. Certainly this
is not the case in the light $Po$ and $Pb$ isotopes. However, for
pedagogical reasons and also to indicate the limitations of the
present approach, we compare the calculated and experimental values in
table 2.

An interesting feature in table 2. is the coupling
between states of different deformations which is not in agreement
with the particle hole picture, see ref. \cite{dup00}. Note also the
influence of triaxiality visible from the drastic change in hindrance
for the decays oblate to prolate in the two different Po isotopes. The
triaxiality parameter $\gamma$ is $-13.9^o$ for ${^{188}Pb}$ and $-3.7$
for ${^{186}Pb}$. The separation of the minima in deformation and
energy obtained from the PES calculations is now the limiting factor
of calculations. Creating a proper collective wave function by means
of the Generator Coordinate Method (GCM) e.g. is a future project that
may enable us to take into account the mixing of the wave functions at
different deformation, resulting in meaningful calculations of all
$\alpha$-decaying shape coexisting nuclei. GCM calculations has been
performed for the mass region, see ref. \cite{dug03}, but has
been restricted to axially deformed states.

To conclude, a simple model to calculate the Hindrance Factors for
alpha decay is presented. Both the energy minima and the hindrance
factor calculations are performed using the same Woods-Saxon potential
with universal parameters. Deformation values are taken from
microscopic calculations and not free parameters. In this
respect, one may view the alpha decay as an important probe to the
mean field wave function. Our calculations show that the HF factors
depend only modestly on deformation changes, as long as these are not
large. However, as soon as the underlying s.p. structure is changed
due to a level crossing e.g., there appears an exponential increase in
the HF with deformation. Our calculations reproduce the experimental
trends for the cases where one deals with clearly separated minima in
the PESs. This indicates that the mean field wave function contain an
essential part of the correlations that are responsible for the
formation of an alpha particle.  In this context it is important to
point out the role of the calculated small deformation values in the
Po- and Rn-region which are normally treated as spherical. Our calculations
also reveal that both protons and neutrons contribute to the HF in
a similar fashion, which is in variance with the simple
shell model like description.

\bibliography{hfaps} 
\bibliographystyle{unsrt}
\end{document}